\UseRawInputEncoding
\documentclass[onecolumn,superscriptaddress]{revtex4-2}

\usepackage{amsmath,amssymb,amsfonts}
\usepackage{graphicx}
\usepackage{algorithm}
\usepackage{algorithmic}
\usepackage{braket}
\usepackage{xcolor}
\usepackage{bm}
\usepackage{subfigure}
\usepackage{appendix}
\usepackage{tikz}
\usetikzlibrary{quantikz2,backgrounds,fit,decorations.pathreplacing}

\begin{document}

\title{Feature Ranking in Credit-Risk with Qudit-Based Networks}

\author{Georgios Maragkopoulos}
\email{giorgosmarag@di.uoa.gr}
\affiliation{Department of Informatics and Telecommunications,
National and Kapodistrian University of Athens, Greece}
\affiliation{Eulambia Advanced Technologies Ltd, Ag. Paraskevi, Greece}

\author{Lazaros Chavatzoglou}
\affiliation{Department of Informatics and Telecommunications,
National and Kapodistrian University of Athens, Greece}

\author{Aikaterini Mandilara}
\affiliation{Department of Informatics and Telecommunications,
National and Kapodistrian University of Athens, Greece}
\affiliation{Eulambia Advanced Technologies Ltd, Ag. Paraskevi, Greece}

\author{Dimitris Syvridis}
\affiliation{Department of Informatics and Telecommunications,
National and Kapodistrian University of Athens, Greece}

\begin{abstract}
In finance, predictive models must balance accuracy and interpretability, particularly in credit risk assessment, where model decisions carry material consequences. We present a quantum neural network (QNN) based on a single qudit, in which both data features and trainable parameters are co-encoded within a unified unitary evolution generated by the full Lie algebra. This design explores the entire Hilbert space while enabling  interpretability through the magnitudes of the learned coefficients. We benchmark our model on a real-world, imbalanced credit-risk dataset from Taiwan. The proposed QNN consistently outperforms LR and reaches the results of random forest models in macro-F1 score while preserving a transparent correspondence between learned parameters and input feature importance. To quantify the interpretability
of the proposed model, we introduce two complementary metrics: (i) edit distance between the model’s feature ranking and that of LR, and (ii) a feature-poisoning test where selected features are replaced with noise. Results indicate that the proposed quantum model achieves competitive performance while offering a tractable path toward interpretable quantum learning.
\end{abstract}

\maketitle

\section{Introduction}

Machine-learning techniques are widely used in contemporary finance for tasks such as credit scoring and fraud detection, where regulatory and institutional requirements demand not only accurate predictions but also transparent, auditable reasoning \cite{Basel}. Linear models like Logistic Regression (LR) meet these demands by providing explicit parameter–feature relationships, yet their simplicity limits their ability to capture the complex, non-linear patterns inherent in financial data. More expressive models—such as deep Neural Networks (NNs) and ensemble methods—can model these complexities effectively but often act as black boxes, making their predictions difficult to interpret \cite{XAI1}. To balance this trade-off, interpretable machine-learning methods have emerged that combine accuracy with transparency; LR  continues to offer coefficient-based clarity, while models like Random Forest (RF) contribute interpretability through feature-importance scores and decision path visualizations, showing that meaningful insights need not come at the cost of explainability \cite{ML1,ML2,ML3}.

Several studies have demonstrated the potential of quantum machine learning (QML) in addressing core financial challenges \cite{QuantumFinanceGeneral}. In credit risk modeling, quantum-enhanced algorithms \cite{QuantumCreditRisk} have shown promise in improving classification performance under noisy and imbalanced data conditions, with some works leveraging quantum kernel methods and hybrid quantum-classical variational models to estimate creditworthiness more effectively than classical baselines. These results suggest that quantum approaches have the potential to uncover complex patterns in financial behavior that may elude conventional techniques.

%In fraud detection, quantum models have similarly begun to exhibit advantages, particularly in detecting subtle or rare fraudulent patterns. Techniques based on variational quantum circuits  have been applied to transaction data, with preliminary results indicating strong performance and potential robustness against adversarial examples \cite{FinancialFraudDetection1,ImbalancedQML1,ImbalancedQML2,FinancialFraudDetection2}. These successes underscore the broader applicability of QML in finance and motivate continued exploration of variational quantum circuits' architectures. 

While most existing works in the emerging field of quantum finance rely on quantum circuits built from qubits (i.e., 2-level quantum systems), in this paper we explore an alternative approach inspired by recent experimental advances and theoretical developments on d-level systems, or qudits~\cite{Wach,Qutrit,Qudit}. Specifically, we employ a quantum neural network (QNN) implemented on a single qudit as our learning model.
Although the Hilbert space of a qudit can be emulated using multiple qubits—$\log_2d$ qubits—the direct use of a single qudit offers a simpler circuit architecture, eliminating the need to distinguish between entangling and non-entangling operations. Furthermore, by adopting an architecture that integrates  encoding and variational layers in a single generating Hamiltonian, we construct a compact QNN in which the trainable weights appear as linear coefficients of the input features. Finally, the inclusion of multiple layers enhances the non-linear representational power of the network and allows for effective cross-talk between features.

We evaluate with the help of standard metrics the performance of the qudit-based QNN in terms of both predictive accuracy and interpretability in terms of feature selection. Specifically, we consider a dataset related to credit risk analysis from Taiwan. For benchmarking purposes, we compare the results with those obtained using standard classical machine learning methods.
To further assess interpretability, we introduce a data ``poisoning'' test in which selected input features are replaced with random noise. A model with strong interpretability is expected to assign low weights to these corrupted features—and our proposed QNN model successfully satisfies this criterion in most cases.

\noindent\textbf{Summary of Contributions}
\begin{itemize}
  \setlength\itemsep{0.25em}
  \item We present a resource-efficient, low-depth qudit QNN architecture that is compatible with current experimental capabilities.
  
  \item The model enables internal feature importance assessment via learned weight magnitudes, eliminating the need for external explanation tools.
  
  \item We introduce a feature-poisoning evaluation method that yields Weighted Interpretability Scores, as we name these, for quantitative attribution analysis.
  
  \item We provide empirical validation on a public large-scale financial dataset, demonstrating competitive or superior predictive performance compared to classical baselines, while maintaining interpretability.
\end{itemize}

The findings of this work, together with those of previous studies such as~\cite{QML4,QML5}, highlight the representational richness of parameterized quantum encodings and help lay the foundation for a broader theory of interpretable quantum learning. In particular, our results demonstrate that compact, fully quantum architectures can be both expressive and explainable—a direction we believe is crucial for the practical deployment of QML in high-stakes domains such as finance.

\section{Background}

\subsection{Qudits}

A qubit describes the state of a two-dimensional quantum degree of freedom, while a qudit generalizes this concept to a $d$-dimensional system. More specifically, a qudit state resides in a $d$-dimensional Hilbert space spanned by the eigenstates of the system's Hamiltonian. Let us denote this orthonormal basis as $\left\{\left|k\right\rangle\right\}_{k=0}^{d-1}$ and let it serve as the computational basis of the qudit. Then, a general pure qudit state can be expressed as
\begin{equation}
\left|\psi\right\rangle = \sum_{k=0}^{d-1} c_k \left|k\right\rangle, \label{basis}
\end{equation}
where the complex coefficients $c_k$ are constrained by the normalization condition $\sum_{k=0}^{d-1} |c_k|^2 = 1$.

In this work, we consider the full $su(d)$ algebra associated with the qudit system, spanned by $d^2 - 1$ traceless Hermitian generators $\{\hat{G}_i\}$. These operators can be chosen to be orthogonal with respect to the Hilbert–Schmidt inner product, such that
\begin{equation}
\mathrm{Tr}\left(\hat{G}_i^\dagger \hat{G}_j\right) = \alpha\, \delta_{i,j},
\end{equation}
with $\alpha$ a positive normalization constant. For $d = 2$, the generators correspond to the Pauli matrices, while for $d = 3$, to the Gell-Mann matrices.

Qudits have attracted increasing attention in recent years \cite{Photonics}, motivated by promising theoretical results \cite{Cozzolino, Wang} in quantum information, quantum computing, and QML. These advances are now driving experimental efforts to realize qudits across a variety of physical platforms, including multi-level Rydberg atoms \cite{Weggemans, Deller}, superconducting circuits \cite{Blok}, trapped ions \cite{Ringbauer} and—most notably—integrated quantum photonic circuits \cite{Kues, Luo,Lapkiewicz}. In photonic implementations, a qudit is encoded in the state of a single photon distributed across $d$ distinct modes, which may correspond to spatial paths, time bins, frequency channels, or orbital angular momentum states.

\subsection{QNN}
Although there are various interpretations of what constitutes a QNN, in recent years the term has increasingly referred to the use of Variational Quantum Circuits (VQCs) for processing classical data. In these approaches, classical data are typically embedded into the quantum circuit—often multiple times—and processed through parameterized quantum gates.

A wide range of VQC architectures has been proposed, differing in how data are encoded and where variational gates are inserted. However, there is currently no clear guideline \cite{Schnabel} on how to tailor these circuits and embedding strategies to exploit potential quantum advantages for specific datasets.

Despite these architectural differences, most QNN share two fundamental features: (i) the extraction of information through projective measurements on the final qubits (or qudit) state, and (ii) the optimization of circuit parameters using a classical optimization algorithm. For this reason, QNN are typically considered hybrid quantum-classical systems—where data are processed within the quantum Hilbert space, but learning and parameter tuning rely on classical computational resources.

In this work, we assume no prior information about the large datasets under consideration. Consequently, we opt for a compiling universal parameterized quantum circuit capable of exploring the full Hilbert space of the quantum processor. To this end, we use a single qudit and the complete set of $d^2 - 1$ generators ${\hat{G}_i}$ of the $su(d)$ algebra. 
We further choose to \textit{condense} the embedding of the input feature vector $\mathbf{x}$ and the variational (trainable) part of the circuit, parameterized by weights $\mathbf{w}$, into a single unitary operation:
\begin{equation}
\hat{U}(\mathbf{x}, \mathbf{w}) = \exp\left(-i \sum_j h(x_j w_j) \hat{G}_j \right), \label{layer}
\end{equation}
where $h(\cdot)$ is a classical function that regularizes the input strength.

A key advantage of this formulation, as we discuss in more detail later, is that it enables a transparent and intrinsic notion of feature importance. Specifically, the learned weights $w_j$ provide insight into the relevance of individual features: small-magnitude weights indicate weak or negligible influence on the model’s output, while large-magnitude weights highlight features with strong predictive significance.
Finally, to enhance the model’s expressivity by increasing the number of trainable parameters, we apply the unitary `layer' of Eq.~(\ref{layer}) multiple times with different weights each time, in order to generate the  qudit's final state
\begin{equation}
    \ket{\psi_{\text{out}}} =  \prod_{\ell=1}^{L} \exp\left(-i \sum_{j} h( x_j w^{(\ell)}_j) \hat{G}_j \right)  \ket{0}\label{manylayers}
\end{equation}
where the index $l$ enumerates the different layers.

\subsection{ Validation of the method }

We demonstrate the efficacy of our method with a real-world financial prediction task, namely the Taiwan credit card default dataset \cite{dataset_Taiwan}. This is a credit risk classification problem where the goal is to estimate the likelihood of a customer defaulting on their credit card payment based on their financial and demographic characteristics. The dataset contains 30,000 records, of which 6,626 (approximately 22.1\%) correspond to default cases. Each record includes 23 input attributes that capture various aspects of a customer’s profile, such as credit limit, gender, marital status, and repayment and billing history over the past six months. The data are labeled as binary outcomes, with 0 representing non-default and 1 indicating default.

%The Australian credit dataset, consists of 690 instances and 14 features where all attribute names and values have been anonymized to meaningless symbols in order to preserve data confidentiality. The third task involves fraud detection in banking transactions, a domain characterized by severe class imbalance as it comprises 284,807 transactions with only 0.17\% labeled as fraudulent. Finally, we include a fourth dataset related to cybersecurity, involving the detection of security breaches based on network traffic patterns. This task models the increasing importance of real-time anomaly detection in digital financial infrastructures. 

We compare the performance of our QNN to three classical baselines: LR, RF and NN. LR serves as a transparent, linear benchmark, while RF provides a nonlinear, ensemble-based approach with moderate interpretability due to their decision tree structure. In contrast, the NN represents a purely data-driven, highly expressive nonlinear model used primarily as a performance ceiling for classification accuracy. Although it cannot provide intrinsic feature rankings or parameter-level interpretability, the NN establishes an upper bound for predictive capability under comparable training conditions. Thus, it serves as a valuable reference point for evaluating how closely the proposed QNN can approach state-of-the-art performance while maintaining a transparent and interpretable internal structure. To evaluate the classification accuracy of different models we employ  F1-score. To assess RF and QNN models' interpretability, we compare the classified list of features provided by these models with the one provided by the LR model using the measure of edit distance. Furthermore, to assess the robustness and transparency of our model, we introduce a feature poisoning workflow in which selected input features are systematically replaced with noise. This perturbative test complements the parameter-based attribution method, showing that features with large learned coefficients indeed correspond to those whose corruption most affects performance.

\subsection{Related Work}
QML provides a potential framework for combining model expressivity with interpretability, particularly by using quantum-native structures that are not directly translatable to classical systems. Recent works have explored QML applications in financial tasks, though the extent to which these models integrate quantum properties into the learning process varies. In \cite{QML1}, the authors employ a neutral-atom quantum processor to solve a quadratic binary optimization problem for model selection in ensemble learning. While this highlights the potential of quantum hardware to contribute to financial modeling, the overall architecture remains classical, with the quantum component functioning solely as an optimizer. As a result, the model's feature encoding and interpretability remain grounded in classical mechanisms, without leveraging quantum state representations or dynamics.

A related approach is found in \cite{QML2}, where a hybrid classical–quantum model is applied to financial forecasting. The study introduces diversity-enhancing determinantal point processes and explores quantum-inspired NN components. However, the quantum circuits primarily augment the classical model, and the interpretability is evaluated through external, post hoc techniques. The separation between classical and quantum components in this architecture constrains the role of quantum dynamics in both representation learning and model explanation.

A more direct focus on interpretability is taken in \cite{QML3}, which proposes an adaptation of the LIME framework \cite{LIME} tailored for quantum models. This method acknowledges the difficulty of generating reliable explanations from probabilistic quantum outputs. Nevertheless, it treats interpretability as an external process by applying surrogate models, rather than embedding it within the quantum model itself. As such, there is no inherent link between parameters and input features, and interpretability remains decoupled from the model’s internal structure.

Our approach shares architectural similarities with the hybrid models developed  in \cite{QML4,QML5}, particularly in how classical data is embedded into quantum circuits. In \cite{QML4}, the authors introduce trainable-frequency feature maps and advocate this design as a promising candidate for expressive QML models. Their work emphasizes spectral bias and classification accuracy. In contrast, while our model is also based on co-encoding data and weights, it shifts the focus toward interpretability by establishing a direct correspondence between parameters and input features. Similarly, the method proposed in \cite{QML5} encodes entire inputs as Hermitian matrices for global evolution, but does not provide explicit feature-wise parameterization. Our model refines this approach by assigning each input feature $x_j$ a dedicated trainable weight $w_j ^{(l)}$, with l being the l-th layer, and generator $\hat{G}_j$, enabling fine-grained attribution and interpretability grounded in quantum dynamics.

\section{Methodology}

Our methodology begins with rescaling the input features, followed by remapping the products of these features with trainable weights. These remapped values are then encoded into a single-qudit QNN via Hamiltonian encoding across the multiple layers of the network, Eq.(\ref{manylayers}).  The quantum processing concludes with the projective measurement of qudit's final state $\ket{\psi_{\text{out}}}$ on the computational basis, Eq.(\ref{basis}). This process is repeated multiple times to estimate measurement probabilities with high accuracy. Subsequently, the model updates the weights using stochastic gradient descent with a a loss that incorporates a regularization term that penalizes large weight norms, i.e., ridge regression. Once the validation error improvement saturates, the model is used for inference. At this stage, the learned weights also serve as indicators of feature importance.  In what follows, we detail the procedure and provide the underlying justification, assuming an $n$-dimensional feature vector $\mathbf{x}=\{x_1,\ldots x_n\}$. We also note that the dimension of the qudit $d$ is chosen such that $d^2-1\ge n$.

%\subsection{Data Encoding}

To ensure numerical stability and encourage effective training dynamics, we remap the generator coefficients in Eq.(\ref{layer}) using a bounded nonlinear function $h(\cdot)$. Specifically, instead of applying raw scalar products $x_j w_{j}^{(l)}$ directly, we use a scaled transformation:
\begin{equation}
  h(x_j w_{j}^{(l)})=  2 \arctan(2 x_j w_{j}^{(l)}) \label{remap}
\end{equation}
as suggested in \cite{Remapping}. This bounds the rotation angles while preserving the sign and relative magnitude of the feature-weight interaction. This way, one avoids excessively large rotations that could destabilize training, but also introduces additional nonlinearity that enhances expressivity.

After applying the  $L$ unitary layers  sequentially to the qudit's `ground' state $\ket{0}$, the resulting state $\ket{ \psi_{\text{out}}}$, Eq.(\ref{manylayers}), is measured to reconstruct a probability distribution over the computational basis: 
\begin{equation}
   p_k =|\braket{k | \psi_{\text{out}}}|^2
\end{equation}
%\[
%    p_k = \frac{|\braket{k | \psi_{\text{out}}}|^2}{\sum_{j} |\braket{j | \psi_{\text{out}}}|^2}
%\]

The output probabilities \( p_k \) are then used to compute the classification loss. For the purpose, we adopt the cross-entropy loss between the predicted distribution \( \{p_k\} \) and the true class labels. Additionally, a regularization term is included to penalize large weight magnitudes, leading to the total loss function:
\begin{equation}
\mathcal{L}_{\text{total}} = \mathcal{L}_{\text{CE}} + \lambda \| \mathbf{w} \|^2,
\end{equation}
where \( \mathcal{L}_{\text{CE}} \) is the standard cross-entropy loss and \( \lambda \) controls the strength of regularization. Training proceeds by minimizing this loss with respect to the weights using gradient-based optimization. In our simulations, we employ the Adam optimizer due to its robustness in handling non-stationary gradients and noisy updates.

Since the entire transformation from \( \mathbf{x} \) to the quantum state \( \ket{\psi_{\text{out}}} \) depends continuously on the trainable parameters, gradients can be analytically computed through partial derivatives. However,  parameter-shift rules, connecting the partial derivatives to circuit's observables for our specific qudit-based Hamiltonian encoding are not yet established. Addressing this issue remains an open avenue for future work.

\subsection{Constructing the generators of the group}

The $d^2 - 1$ generators of $\mathfrak{su}(d)$ can be categorized  \cite{Qudits_Bloch} into three distinct sets: symmetric, anti-symmetric and diagonal. The off-diagonal symmetric generators (real) are defined  as
\begin{equation}
\hat{g}^{(R)}_{jk} = |j\rangle\langle k| + |k\rangle\langle j|
\end{equation}
with  $0 \leq j < k \leq d-1$.
These matrices are real, symmetric, and Hermitian. The off-diagonal antisymmetric generators (imaginary) are defined for the same index pairs as
\begin{equation}
\hat{g}^{(I)}_{jk} = -i|j\rangle\langle k| + i|k\rangle\langle j|.
\end{equation}
These matrices are purely imaginary, anti-symmetric, and also Hermitian. The diagonal traceless generators are defined for each $l$ with $0 \leq l \leq d - 2$ as
\begin{equation}
\hat{g}^{(D)}_l = \sqrt{\frac{2}{(l+2)(l+1)}} \left( \sum_{j=0}^{l} |j\rangle\langle j| - (l+1) |l+1\rangle\langle l+1| \right),
\end{equation}
which are Hermitian and traceless by construction. 

Together, these three classes of generators span the $\mathfrak{su}(d)$ algebra. To generate this set computationally, we implement an algorithm that constructs each type of generator directly in matrix form. The algorithm is shown in Algorithm~\ref{alg:su_d_generators}.

\begin{algorithm}[H]
\caption{Construction of \(\mathfrak{su}(d)\) generators}
\label{alg:su_d_generators}
\begin{algorithmic}[1]
\REQUIRE Dimension \(d \in \mathbb{Z}\)
\ENSURE List of \(d^2 - 1\) Hermitian, traceless generators of \(\mathfrak{su}(d)\)

\STATE Initialize empty list \texttt{generators}.

\FOR{\(j = 0\) to \(d-2\)}
    \FOR{\(k = j+1\) to \(d-1\)}
        \STATE Create matrix \(\hat{g}^{(R)}\) with 1 at \((j,k)\) and \((k,j)\)
        \STATE Append \(\hat{g}^{(R)}\) to \texttt{generators}
        \STATE Create matrix \(\hat{g}^{(I)}\) with \(-i\) at \((j,k)\), \(+i\) at \((k,j)\)
        \STATE Append \(\hat{g}^{(I)}\) to \texttt{generators}
    \ENDFOR
\ENDFOR

\FOR{\(l = 1\) to \(d-1\)}
    \STATE Create diagonal matrix \(\hat{g}^{(D)}\) with:
    \begin{itemize}
        \item Value \(\sqrt{\frac{2}{l(l+1)}}\) on diagonal entries \((0,0)\) to \((l-1,l-1)\)
        \item Value \(-l \sqrt{\frac{2}{l(l+1)}}\) on entry \((l,l)\)
    \end{itemize}
    \STATE Append \(\hat{g}^{(D)}\) to \texttt{generators}
\ENDFOR

\RETURN \texttt{generators}
\end{algorithmic}
\end{algorithm}

In practice, the algorithm is implemented in PyTorch to support automatic differentiation and integration with QML frameworks. The resulting set $\{\hat{G}_i\}$ is then used to build parameterized Hamiltonians of the form
\begin{equation}
\hat{H}(\mathbf{x}, \mathbf{w}) = \sum_{j} x_j~w_j \hat{G}_j.
\end{equation}
In what follows we explain how this unified encoding enables an interpretable QNN.

\subsection{Model's interpretability}

\begin{figure*}[t]
  \centering
%    \hspace*{0\textwidth}
    \includegraphics[width=0.78\textwidth]{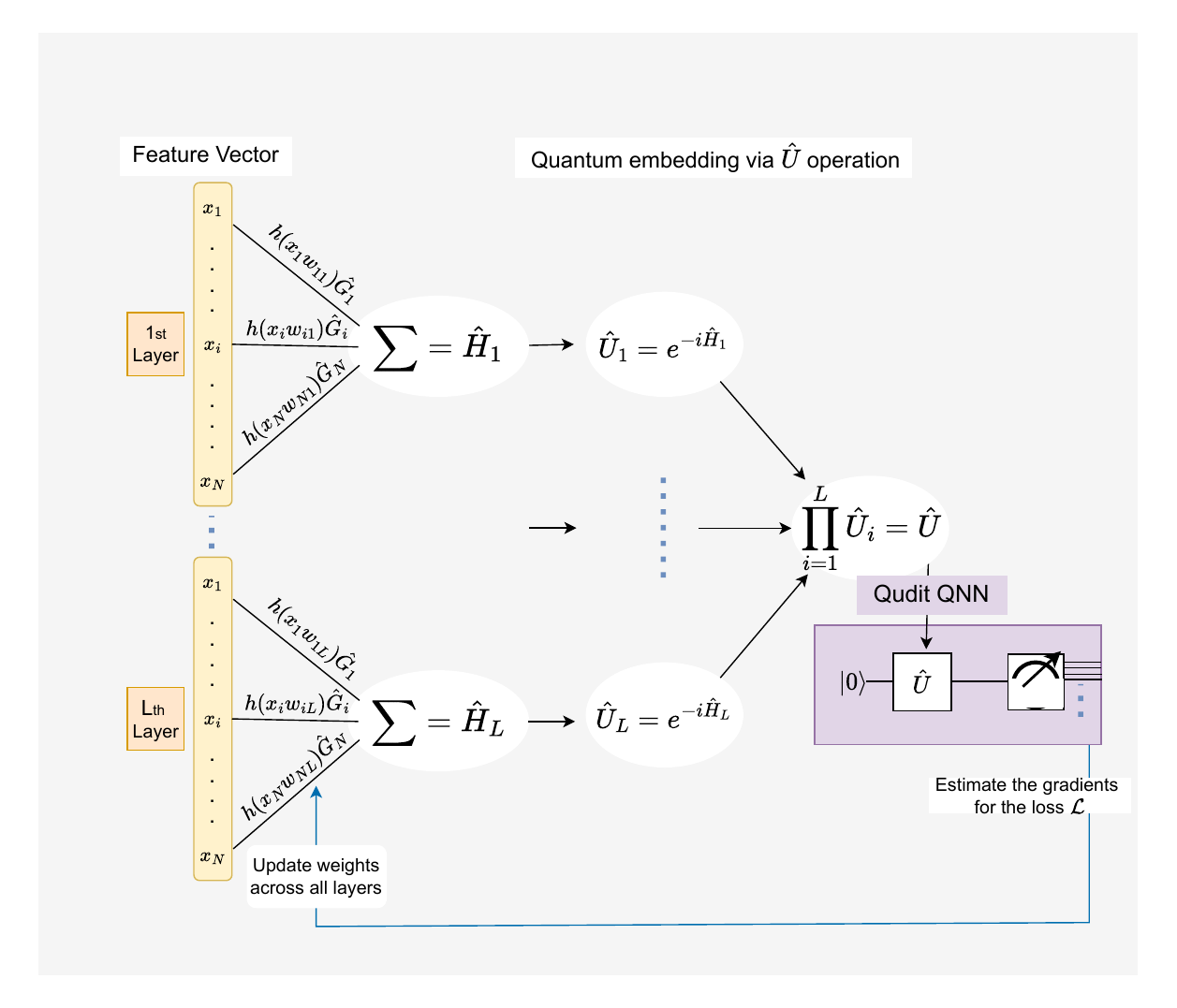}
    \caption{Multilayer QNN architecture with Hamiltonian-based feature encoding. At each layer, classical input features are re-mapped and combined with trainable weights to form coefficients for a set of \(\mathfrak{su}(d)\) generators, defining the layer Hamiltonian. This process is repeated over \(L\) layers in a data re-uploading fashion. The resulting unitary operations sequentially act on a single base qudit, followed by measurement. The model parameters are optimized iteratively using gradient descent.      \label{fig2}}
\end{figure*}

 Our QNN model implements a data-encoding and weight-parameterized unitary operator, as defined in Eq.~(\ref{manylayers}) and illustrated in Fig.~\ref{fig2}. This design contrasts with conventional variational quantum circuits, where data encoding and trainable parameters are typically implemented through separate gates, see for instance \cite{QML4,QML5}. In our approach, both are co-encoded within the same Lie algebraic generator structure, resulting in a circuit that is not only more compact, but up to some degree, analytically interpretable.

Let us consider a single layer circuit as in Eq.(\ref{layer}) and ignore the effect $\arctan$ function $h(\cdot)$ for the moment. The mapping $\mathbf{x} \mapsto \hat{U}(\mathbf{x}, \mathbf{w})\ket{0}$ defines a quantum feature map, where the classical data is embedded into the Hilbert space via a nonlinear transformation induced by the exponentiation of a data-dependent Hermitian operator. Although the Hamiltonian $\hat{H}(\mathbf{x}, \mathbf{w}) = \sum_j x_j w_j \hat{G}_j$ is linear in vector $\mathbf{x}$, the matrix exponential generates a power series \cite{Nielsen_Chuang_2010}:
\begin{equation}
    \hat{U}(\mathbf{x}, \mathbf{w}) = \mathbb{I} - i \hat{H}(\mathbf{x}, \mathbf{w}) - \frac{1}{2!} \hat{H}(\mathbf{x}, \mathbf{w})^2 + \frac{i}{3!} \hat{H}(\mathbf{x}, \mathbf{w})^3 + \cdots
\end{equation}
Since the generators $\{\hat{G}_j\}$  do not commute, the powers of $\hat{H}(\mathbf{x}, \mathbf{w})$ in this expansion give rise to nonlinear interaction terms between features. That is, even though the input enters linearly, the unitary evolution encodes higher-order and mixed feature terms implicitly. 

From an interpretability perspective, the model assigns each input feature $x_j$ to a generator $\hat{G}_j$, modulated by a learned parameter $w_j$. When $x_j w_j \approx 0$, the contribution of $\hat{G}_j$ to the overall unitary $\hat{U}(\mathbf{x}, \mathbf{w})$ is negligible, effectively freezing that direction in Hilbert space. As a result, features associated with low-magnitude weights are interpreted as uninformative. Conversely, large values of $|x_j w_j|$ induce stronger rotations in Hilbert space around the axis defined by $\hat{G}_j$, signaling higher feature relevance. This mechanism underpins the model's feature attribution capability.

Proceeding now with the multiple layer Eq.(\ref{manylayers}) that we are employing for our numerical tests. 
This architecture resembles deep classical networks, but with each layer operating in the quantum state space. Each layer contributes additional expressivity, and their compositional nature enables modeling of more complex decision boundaries. The output of the final layer is projected onto a measurement basis to generate the prediction \cite{QNN_general}.

In the multilayer case, the total unitary is a product of exponentials:
\begin{equation}
\hat{U}_{\text{total}} = \prod_{\ell=1}^{L} \exp\left(-i \sum_j x_j w_j^{(\ell)} \hat{G}_j\right)
\end{equation}
To understand how feature influence accumulates, consider two layers and apply the Baker–Campbell–Hausdorff (BCH) formula:
\begin{equation}
e^{\hat{A}} e^{\hat{B}} = \exp\left(\hat{A} + \hat{B} + \frac{1}{2}[A,\hat{B}] + \cdots\right)
\end{equation}
where
\begin{equation}
\hat{A} = -i \sum_j x_j w_j^{(1)} \hat{G}_j, \quad \hat{B} = -i \sum_j x_j w_j^{(2)} \hat{G}_j.
\end{equation}
The first-order term simplifies to:
\begin{equation}
\hat{A} + \hat{B} = -i \sum_j x_j \hat{G}_j (w_j^{(1)} + w_j^{(2)}),
\end{equation}
showing that weights across layers add up linearly. Thus, features with consistently small weights across layers have negligible impact on the evolution across every order of the expansion, while features with large cumulative weights contribute strongly in the first-order linear part. Higher-order commutator terms introduce nonlinear feature interactions, enhancing expressivity without obscuring the leading-order interpretability.

Because the transformation is controlled by the product $x_j w_j$, the magnitude of the learned parameter  $w_j$ modulates the influence of feature $x_j$ on the quantum evolution. Thus, after training, features associated with high-magnitude weights are considered more important. In contrast, weights that converge to values close to zero correspond to generators that produce negligible evolution under the given input, suggesting that the associated features do not meaningfully affect the model’s behavior. 
%We empirically validate this hypothesis through a feature poisoning experiment.

\subsection{Metrics for quantifying model's interpretability }

Assuming the LR model provides the ground-truth ranked list of features, the interpretability of the QNN and RF models can be estimated by computing the edit distance between this reference list and the feature-importance rankings generated by each model. The edit distance represents the minimum number of insertions, deletions, or swaps required to transform one ranking into another, thereby serving as a direct measure of alignment between the QNN or RF and the interpretable baseline (LR). A smaller edit distance indicates greater alignment in feature importance—and hence higher interpretability consistency—with the LR model.

To further examine interpretability and model robustness, we conducted a poisoning-based evaluation in which selected input features were replaced with Gaussian noise during inference. This experiment assesses how each model’s performance degrades when informative features are corrupted. In more detail, for each model under study, we compute the Weighted Interpretability Score (WIS) which weights the contribution of each feature by its assigned importance. 
In practice, the computation proceeds by comparing the model’s top-ranked features with the set of truly informative ones identified from the clean dataset. For each feature, its importance weight is added to the overall score if it belongs to the true informative set, and subtracted if it does not. The resulting value captures the net alignment between the model’s learned feature importances and the actual predictive variables. A higher WIS therefore reflects a model that concentrates its learned weights on genuinely relevant inputs while assigning little or no importance to noisy or spurious features.

%Formally:
%\[
%\text{WIS} = \sum_{f \in R_k \cap T} I(f) - \sum_{f \in R_k \setminus T} %I(f),
%\]
%where \(T\) is the ground-truth set of informative features, \(R_k\) the top-\(k\) features ranked by the model, and \(I(f)\) denotes feature importance. 

\section{Numerical Results}

This study presents a numerical comparison of the performance of three classical models --LR, RF, and NN-- against the proposed qudit-based QNN on a real-world credit risk prediction problem. Each model was trained multiple times with different random seeds to account for variability in initialization and feature selection. Furthermore, to assess the effect of model complexity, we evaluated several sizes of the RF, NN, and qudit QNN models, each corresponding to a different number of trainable parameters. All experiments were conducted on the Taiwan Credit Risk dataset, which comprises $30,000$ customer profiles with $23$ financial features and a binary target variable indicating default or non-default. The qudit dimension was set to $d=5$, and the implementation code for the QNN model is publicly available in \cite{GitHub}. Each model was ran for 10 iterations with different random seeds each time, in order to collect statistical results.

We report results for two complementary evaluation settings:  
\textit{(i)} training and testing on the uncorrupted dataset, which enables comparison using the \textit{edit distance} interpretability metric; and  
\textit{(ii)} training on clean data but testing with feature corruption (data poisoning), used to assess robustness and interpretability via the WIS metric. We evaluate classification accuracy (F1 score) for both  uncorrupted and corrupted with noise,  datasets. The second case,  provides information on the robustness  of the predictive performance for each model. 

\subsection{Evaluation of Performance via the Edit Distance}

In the first setting, all models were trained and evaluated on the unmodified dataset. The goal was to compare predictive accuracy and interpretability under normal operating conditions.
Since NNs do not provide a consistent feature ranking, the edit distance was computed only for the QNN and RF models, using the feature ordering obtained from the LR coefficients as the reference. 

\begin{figure*}[t]
    \centering
    \includegraphics[width=0.95\textwidth]{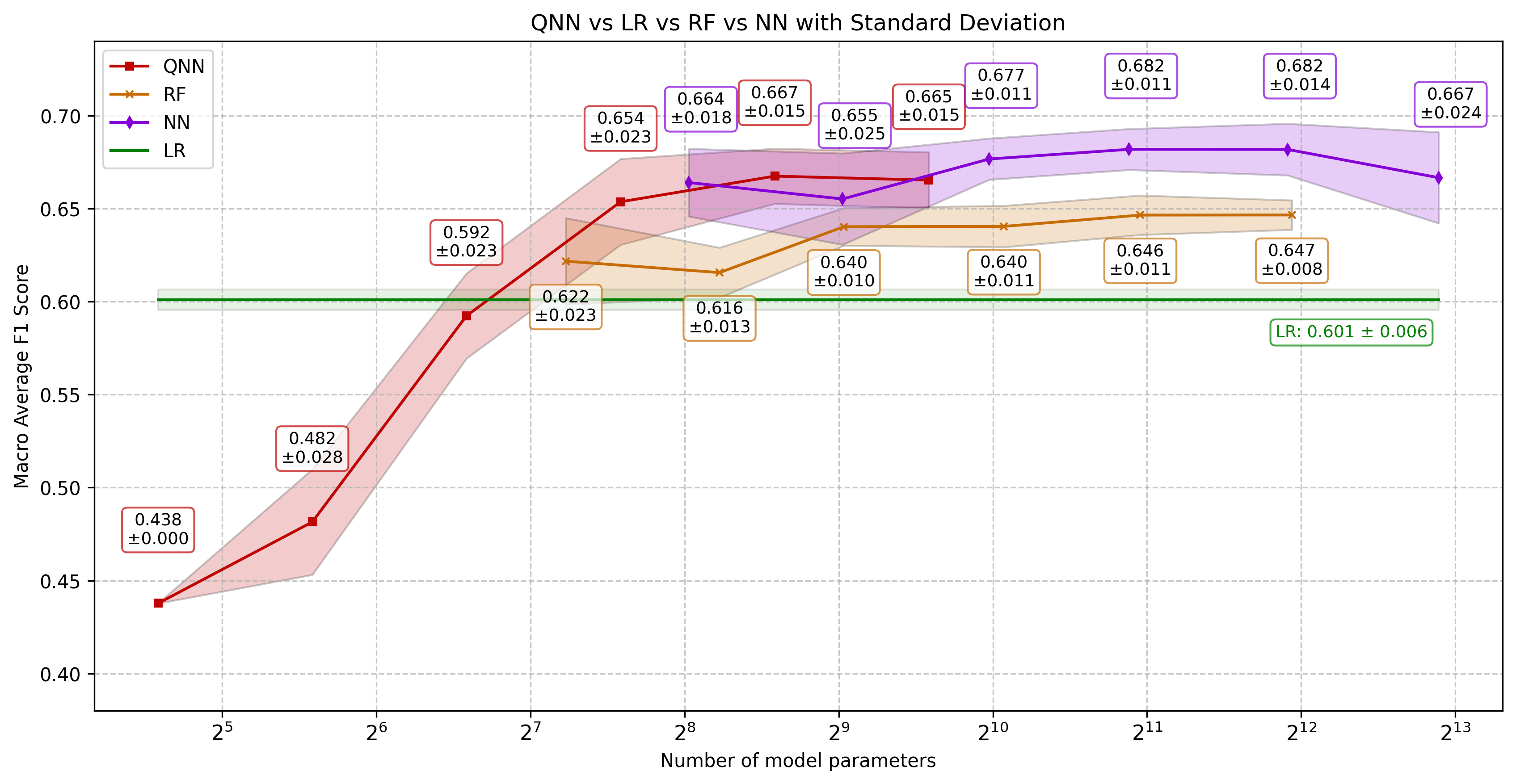}
    \caption{Macro-averaged F1-score of the QNN, RF, NN, and LR on the Taiwan credit dataset. Shaded regions denote standard deviations across runs.}
    \label{fig:f1_full}
\end{figure*}

\begin{table}[h]
\centering
\begin{tabular}{|l|c|c|c|}
\hline
\textbf{Model} & \textbf{\# Params} & \textbf{F1 score} & \textbf{Edit Distance to LR} \\
\hline
LR & -- & 0.6010 $\pm$ 0.006 & -- \\
RF & 3927 & 0.647 $\pm$ 0.008 & 21.10 $\pm$ 0.99 \\
NN & 1853 & 0.682 $\pm$ 0.011 & -- \\
\textbf{Qudit QNN} & \textbf{384} & 0.667 $\pm$ 0.015 & \textbf{20.9 $\pm$ 1.85} \\
\hline
\end{tabular}
\caption{Performance and interpretability on the  Taiwan dataset. The models presented are the ones with the highest performance in F1 score. Lower edit distance indicates stronger alignment with LR’s feature ordering.}
\label{tab:full_data}
\end{table}

\begin{figure}[h]
    \centering
    \includegraphics[width=0.48\textwidth]{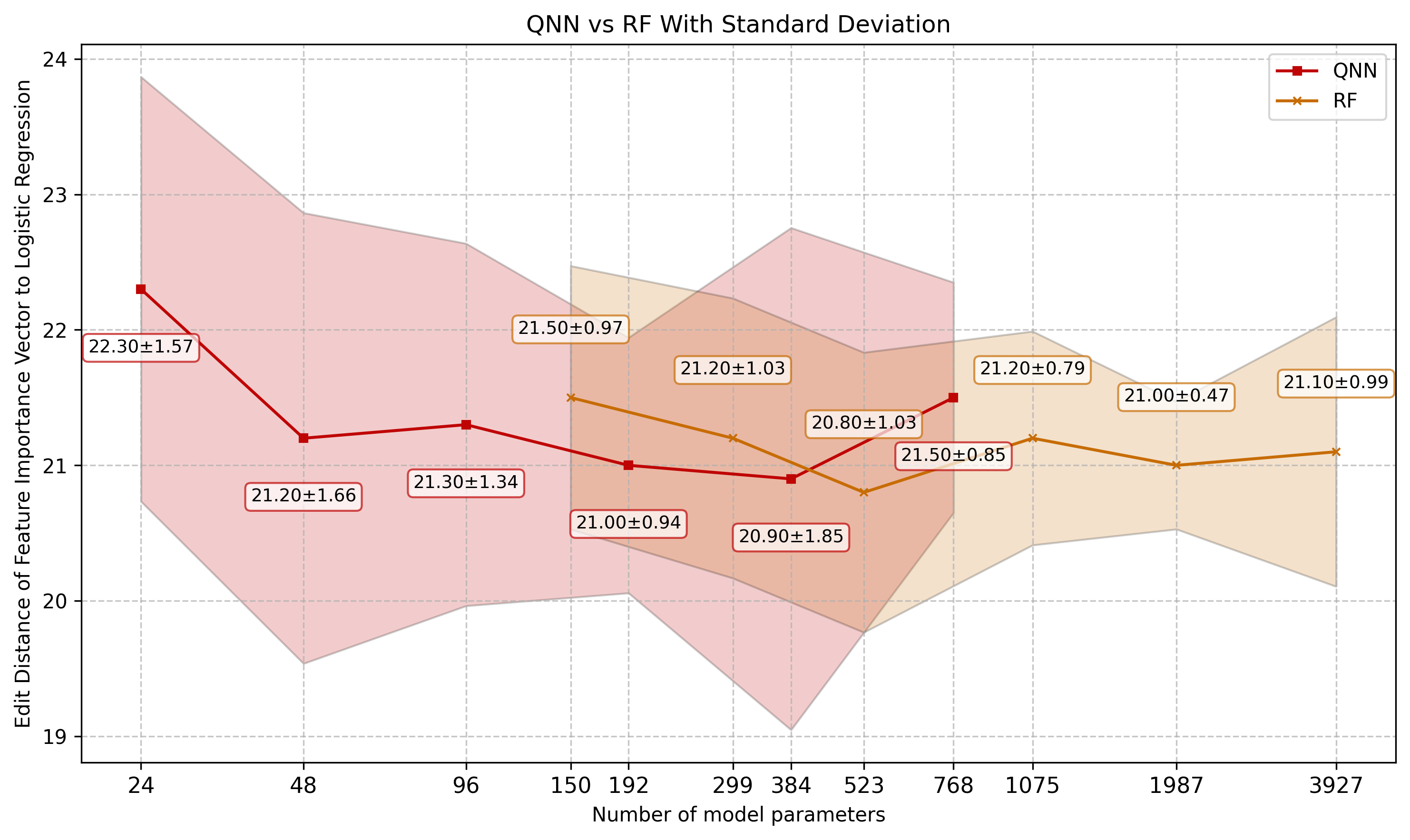}
    \caption{Edit distance between feature-importance vectors of QNN and RF relative to LR as a function of model size. Lower values indicate closer alignment with LR’s interpretability.}
    \label{fig:edit_distance}
\end{figure}

In Fig.~\ref{fig:f1_full} we present the macro average F1 score achieved by the models under study
for different number of trainable parameters. Then for each model we select the highest accuracy score that is  presented in  Table~\ref{tab:full_data} together with the related interprtability as evaluated by the edit distance. 
As shown in Table~\ref{tab:full_data}, the QNN achieves a macro-averaged F1-score of approximately 0.667, outperforming LR and closely matching the classical NN baseline. Importantly, there is no statistical difference in the  QNN’s edit distance to LR, compared with RF's, suggesting that it captures similar feature relevance patterns while maintaining higher overall predictive accuracy. The F1-curve of the Qudit QNN in Fig.~\ref{fig:f1_full} consistently lies above that of the RF across all evaluated configurations, confirming that the proposed quantum model surpasses conventional ensemble methods in predictive performance while maintaining a fraction of their parameter count. Furthermore, the NN’s curve plateaus slightly higher than the QNN’s, indicating that classical deep learning still provides marginally superior accuracy, albeit at the expense of interpretability.

Fig.~\ref{fig:edit_distance} complements the analysis by illustrating the \textit{edit distance} between the feature-importance rankings of the QNN and RF relative to the interpretable LR baseline. Both quantum and classical ensemble models achieve edit distances around 21, suggesting comparable alignment with the feature relevance patterns identified by LR. Importantly, the QNN maintains this interpretability while requiring significantly fewer parameters (384 vs. 3927 for RF), demonstrating a favorable trade-off between expressivity, interpretability, and resource efficiency. This indicates that the quantum model can reconcile interpretability with expressive, nonlinear learning.

\subsection{Performance using poisoning evaluation}

To evaluate model robustness and interpretability under imperfect input conditions, we performed a controlled feature-poisoning experiment. In each evaluation run we randomly selected seven features and replaced their values with Gaussian noise; the selection was independent and uniformly random across iterations. This procedure creates a clear, reproducible split between genuine features and corrupted (noise-replaced) features, providing a direct, ground-truth testbed for interpretability and robustness.

\begin{table}[h]
\centering
\begin{tabular}{|l|c|c|c|}
\hline
\textbf{Model} &  \textbf{\# Params} & \textbf{F1 (Poisoned Data)} & \textbf{WIS} \\
\hline
LR &  - &0.579 $\pm$ 0.020 & 0.932 $\pm$ 0.043 \\
RF &   3927   & 0.614 $\pm$ 0.027 & 0.953 $\pm$ 0.026 \\
\textbf{Qudit QNN}  & 384 & \textbf{0.632 $\pm$ 0.040} & 0.853 $\pm$ 0.048 \\
\hline
\end{tabular}
\caption{Performance and interpretability under feature corruption (poisoning). Higher WIS values indicate better identification of truly informative features.}
\label{tab:poisoning}
\end{table}

As summarized in Table~\ref{tab:poisoning}, the Qudit QNN achieves the highest classification performance under feature corruption, maintaining an average F1-score of 0.632 despite noise injection. This indicates that the quantum model remains stable even when key inputs are perturbed, outperforming both the LR (0.579) and RF (0.614) baselines in predictive robustness.

In terms of interpretability, the RF preserves the strongest alignment with the LR reference, achieving the highest BIS and WIS values (0.844 and 0.953, respectively). The QNN, while slightly lower in interpretability (BIS = 0.606, WIS = 0.853), maintains a meaningful correspondence between its learned parameters and the true informative features. This suggests that the model does not overfit spurious correlations, even in the presence of corrupted inputs.

Overall, these results highlight a clear trade-off: the Qudit QNN offers superior robustness and competitive accuracy while preserving moderate interpretability, whereas the RF retains stronger interpretability but at the cost of reduced resilience to noise. This balance between predictive power and transparency positions the QNN as a promising architecture for reliable and interpretable credit-risk modeling under uncertain or noisy data conditions.

\subsection{Discussion}

The experiments reveal that while classical NN achieve slightly higher F1-scores on the uncorrupted dataset, they lack an intrinsic mechanism for feature ranking. In contrast, the qudit QNN achieves a balance between performance and interpretability: its F1-score remains close to that of NN, while its feature importance aligns more closely with LR than that of RF. When combined with its robustness to data corruption, these findings suggest that the QNN provides a meaningful step toward interpretable and reliable quantum learning models applicable to real-world financial decision-making.

\section{Conclusion}
This study presented a qudit-based QNN employing Hamiltonian co-encoding of data and trainable parameters within the $su(d)$ algebra. Evaluations on the Taiwan credit-risk dataset demonstrate that the QNN achieves comparable predictive performance to classical NNs while maintaining interpretability comparable to RF.

Results further show that increasing the number of layers systematically enhances classification performance, suggesting that deeper quantum architectures leverage the model’s representational capacity more effectively. Importantly, the proposed encoding strategy remains one of the few quantum formulations in the current literature to achieve high accuracy on a real-world financial dataset.

Overall, the qudit-based QNN offers a promising balance between model transparency and predictive strength, positioning it as a viable candidate for interpretable QML in finance. Future work will focus on extending the architecture to multi-qudit implementations and exploring applications where accountability and interpretability are paramount.

\section*{Acknowledgments}
This work was supported by the European Union’s Horizon Europe research and innovation program under grant agreement No.101092766 (ALLEGRO Project) and the Hellas QCI project under the Digital Europe Programme, grant agreement No.101091504.

\bibliographystyle{apsrev4-2}
\bibliography{bibliography}

\end{document}